\documentclass[10pt]{article}

\usepackage{calrsfs}
\usepackage{amsmath,amsfonts,latexsym}
\usepackage{amssymb}
\newtheorem{Definition}{Definition}
\newtheorem{lemma}{Lemma}
\newtheorem{theorem}{Theorem}

\newtheorem{proposition}{Proposition}
\newtheorem{corollary}{Corollary}
\usepackage{graphicx}
\graphicspath{%
    {converted_graphics/}
    {ARXIVfigures/}
}
\begin{document}

\title{The Structure of Optimal and Near Optimal Target Sets in Consensus Models}   
\author{Fern Y. Hunt\\
Applied and Computational Mathematics Division\\
National Institute of Standards and Technology\\
Gaithersburg,Maryland 20899}        
\date{\today}          
\maketitle
%
%
\begin{abstract}
We consider the problem of identifying a subset of nodes in a network that will enable the fastest spread of information in a decentralized environment.In a model of communication based on a random walk on an undirected graph, the optimal set over all sets of the same or smaller cardinality minimizes the sum of the mean first arrival times to the set by walkers starting at nodes outside the set. The problem originates from the study of the spread of information or consensus in a network and was introduced in this form by V.Borkar et al. in 2010.  More generally, the work of A. Clark et al. in  2012 showed that estimating the fastest rate of convergence to consensus of  so-called leader follower systems leads to a consideration of the same optimization problem. 

The set function $F$ to be minimized is supermodular and therefore the greedy algorithm is commonly used to construct optimal sets or their approximations. In this paper, the problem is reformulated so that the search for solutions is restricted to optimal and near optimal subsets of the graph. We prove sufficient conditions for the existence of a greedoid structure that contains feasible optimal and near optimal sets. It is therefore possible we conjecture, to search for optimal or near optimal sets by local moves in a stepwise manner to obtain near optimal sets that are better approximations than the factor $(1-1/e)$ degree of optimality guaranteed by the use of the greedy algorithm. A simple example illustrates aspects of the method.
\end{abstract}
%
%
%
%
\section{Random Walk Consensus Model}\label{S:rwalk}      

Given a connected graph $G=(V,E)$, with vertices or nodes $V$ and edges $E$, we imagine a
random walker situated at a node $i \in V$, moving to another node $j \in V$ in a single discrete time
step. The choice of $j$ is random and has probability, 
%
\begin{equation}\label{E:probtrans}
Prob\{i \rightarrow j\}=
\begin{cases}
p_{i j}>0 , &\text{if   }  (i,j) \in E \\
p_{i j}=0    &\text{otherwise.}
\end{cases}
\end{equation}
The matrix $\mathbb{P}=(p_{ij})_{i,j=1 \cdots N}$ is the transition matrix of a Markov chain which in
this paper, is assumed to be irreducible and aperiodic (\cite{Kemeny}).  $N$ is the number of nodes and
as in \cite{Borkar} the spread of information is described in terms of a process that is dual to
the movement from informed to uninformed nodes. A random walk begins  outside
a pre-determined set $A$ of informed target nodes and ends at $A$. Starting at node $i \notin A$, a random walker first reaches the
set $A$ at a hitting time $T_{A}=\min\{n >0: X_{n} \in A \}$, where $X_n$ is the node occupied by the walker
at time $n$. The hitting time is closely related to the rate of  convergence in a leader-follower model of Clark et al., as well as other consensus models \cite{Clark, Barahona}, \cite{Borkar}. The effectiveness of a set A in the spread of information by random walks can then be measured by,
\begin{equation} \label{E:hitnumber}
F(A)=\left({\sum}_{i \notin A}h(i,A)\right),
\end{equation}
where $h(i,A)=\mathbf{E_{i}}[T_{A}]$, is the expected number of steps to $A$ starting at node $i$. When
$F(A)$ is small, $A$ is a desirable choice of informed nodes, but is a poor choice if $F(A)$ is large. A 
standard result in Markov chain theory tells us that $h(i,A)$ is the ith component of the vector $\mathsf{H}$,
which solves the linear equation.
%
\begin{equation}\label{E:lineq}
{\LARGE\mathsf{H}}=\bf{1}+\mathbb{P}_{A}{\LARGE\mathsf{H}}
\end{equation}
where $\mathbf{1}$ is a column vector of $N-|A|$ ones and  $\mathbb{P}_{A}$ is the matrix that results from
crossing out the rows and columns of $\mathbb{P}$ corresponding to the nodes of $A$ \cite{Kemeny}.
Limited resources can constrain the maximum size of the subset to be selected so it makes sense then
to ask for the most effective spreader subject to a cardinality constraint, i.e.
%
\begin{equation}\label{E:opt}
\min_{ A \subset V,\\\ |A|\leq M} F(A)
\end{equation}


 Borkar et al. \cite{Borkar} showed that for arbitrary subsets
$A,\ B \subseteq V \,$,  $F(A\cap B) +F(A\cup B)\geq F(A)+F(B)$, that is, $F$ is a supermodular function.  Clark et al. in \cite{Clark} discussed a continuous time leader-follower problem where a set $A$ of leader nodes are assigned fixed function values and the remaining follower nodes update their function values by weighted exchange with their neighbors with weights defined by equation (\ref{E:probtrans}). If $G$ is strongly connected, the node function values converge to a consensus value (vector) determined by the leader nodes and the rate of convergence has a connection to precisely the random walk problem we are describing.  In \cite{Clark}, two optimization problems are posed. The first, is to select up to $M$ leaders in order to minimize the convergence error. Given the random walk connection the problem in Clark is 
equivalent to the the problem posed in equation (\ref{E:opt}).

Since $F$ is supermodular, both references \cite{Borkar},\cite{Clark} make use of the work of \cite{NemWolFish} to devise 
a greedy algorithm that builds an approximate solution to the optimization problem (\ref{E:opt}) in a stepwise fashion until a set of 
cardinality $M$ is reached. Recall that at the first stage of construction, the node with the smallest value of $F$ among the nodes is selected. At the $p$th stage the node is added to the set that results in a set of cardinality $p+1$ with the smallest value of $F$.  Using the results of \cite{NemWolFish}, Borkar et al. were able to give some guarantee of the 
quality of the approximation of a weaker version of the optimization problem with the additional constraint that the set 
contain an element $\{a\}$. If $F_{a}^{*}$ is the optimal value of this weaker optimization problem and $A_M(a)$ is the greedy 
approximation obtained by starting the algorithm with the set $\{a\}$, then
%
\begin{equation} \label{E:upperbound}
 F(\{a\})-F(A_{M}(a))\geq (1-\frac{1}{e})(F(\{a\})-F_{a}^{*})
\end{equation}
In the continuous time setting, Clark et al. obtained a similar inequality but it is independent of the choice of a required initial
element. \\

It will be instructive in what follows to see the results of applying the greedy algorithm to a specific graph. In all the
examples discussed in this paper we assume that for every neighbor $j$ of node $i$, $p(i,j)=\frac{1}{deg(i)}$. \\ \\

\begin{figure}[tbp] 
  \centering
  \includegraphics[width=5.02in,height=3.23in,keepaspectratio]{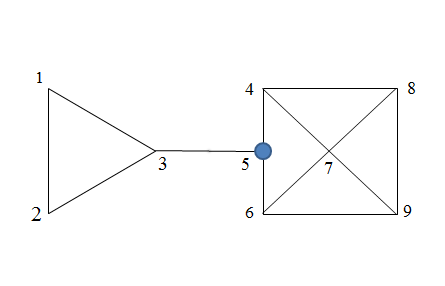}
  \caption{graph for EXAMPLE 1 with 9 vertices showing optimal K=1 set}
  \label{fig:JW8K1}
\end{figure}

\begin{figure}[tbp] 
  \centering
  \includegraphics[width=5.02in,height=3.59in,keepaspectratio]{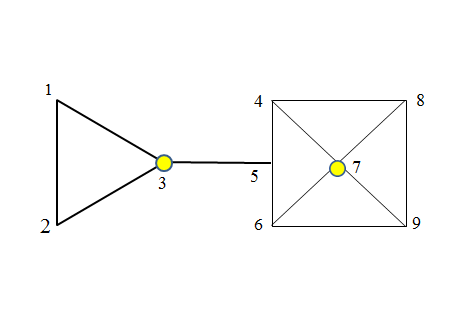}
  \caption{graph for EXAMPLE 1 with 9 vertices showing optimal K=2 set}
  \label{fig:JW8K2}
\end{figure}

\begin{figure}[tbp] 
  \centering
  \includegraphics[width=5.67in,height=3.69in,keepaspectratio]{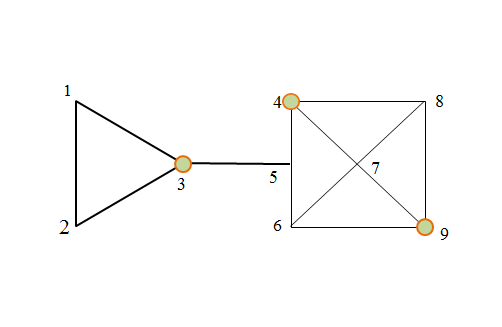}
  \caption{graph for EXAMPLE 1 with 9 vertices showing optimal K=3 set}
  \label{fig:JW8K3}
\end{figure}

\begin{figure}[tbp] 
  \centering
  \includegraphics[width=4.93in,height=3.64in,keepaspectratio]{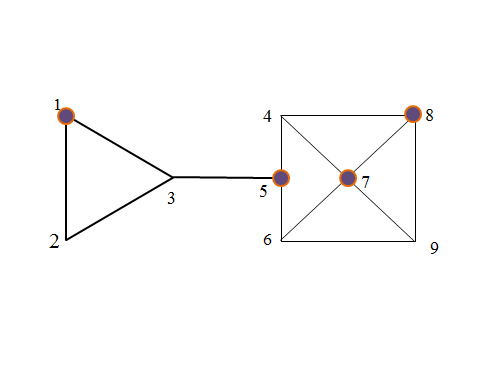}
  \caption{graph for EXAMPLE 1 with 9 vertices showing optimal K=4 set }
  \label{fig:JW8K4}
\end{figure}

\begin{figure}[tbp] 
  \centering
  \includegraphics[width=5.67in,height=3.35in,keepaspectratio]{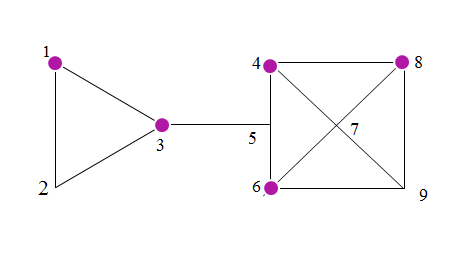}
  \caption{graph for EXAMPLE 1 with 9 vertices showing optimal K=5 set}
  \label{fig:JW8K5}
\end{figure}

{\bf EXAMPLE 1}
Figures \ref{fig:JW8K1}-\ref{fig:JW8K5} show the picture of a graph $G$ with $9$ nodes including the nodes of the optimal set of cardinality $K$ for $K=1$ through $5$. In this example the optimal set of a given cardinality does not contain an optimal subset that is one element smaller.
Thus, for example the optimal $3$ element set cannot be obtained from the (unique) $2$ element set. In turn the $2$ element set does not contain the $1$ element optimal set. Therefore the greedy algorithm does not produce optimal sets.\\ \\

In this paper,the optimization problem is reformulated with a view towards improving the guarantees of the greedy algorithm  proved in \cite{Borkar}, \cite {Clark}. We seek solutions or approximations of the problem for some cardinality $M$ where $M <K$ where $K$ fixed, is the cardinality of a vertex cover of the graph. A vertex cover is an optimal set for its cardinality, so we first seek solutions of problem (\ref{E:opt}) that are subsets of the cover. This is done in section \ref{S:newsolution} in order to provide motivation for our method. Since (as we show) it is not always possible to find optimal subsets in an arbitrary vertex cover, the search space is enlarged to a class of optimal and near optimal sets of a specified degree of optimality relative to the vertex cover (see the definition in (\ref{E:Lck})). These sets are introduced in section \ref{S:optnopt}. Our main result is a sufficient condition for the existence of a gredoid structure containing optimal and near optimal sets (see Theorem \ref{T:structure}
 and Corollaries \ref{C:case1} and \ref{C:case2}). The properties of the greedoid enable one to make local moves (for example by adding or deleting certain elements of a set) that preserve the optimality or near optimality of feasible sets. We have observed that
solutions of the original optimization problem are well approximated by high ranking near optimal sets and the quality of the approximations can exceed the $(1-1/e)$ factor guaranteed by the greedy algorithm. In section \ref{S:method} we show an illustrative example. The key is the ability to improve the approximations by searching the greedoid structure. This is formalized in 
section \ref{S:graph}, where we briefly introduce the greedoid graph whose nodes are feasible sets. Properties of adjacent nodes in this graph enable local moves among sets that can be used to improve approximations and thus lay the groundwork for deterministic local search methods such as branch and bound on the one hand or stochastic search on the other. The paper concludes in section \ref{S:concl} with a summary of the results and questions for further research.
%
\section{Finding and Approximating Optimal Sets} \label{S:newsolution}
\subsection{Maximal Matches}\label{S:maxmatch}

The optimization problem as posed in equation(\ref{E:opt}) assumes no advance knowledge about the optimal set or 
other possibly related subsets of $G$. In this section we seek to explore alternative formulations of the problem that could lead
to better approximations of the optimal set. The next definition will be helpful in the discussion that follows:
\begin{Definition}
A \underline{vertex cover} of a graph $G=(V,E)$ is a set of vertices that are incident to every edge in $E$.
\end{Definition}
\begin{lemma}\label{VCopt}
Let $A$ be a vertex cover of the graph $G$, with $K=|A|$. Then $A$ solves the optimization problem (\ref{E:opt}) for $M=K$.
\end{lemma}
{\bf Proof:}
Since every edge of $G$ is incident to an element of $A$, a random walker starting at a vertex $i$
outside of $A$ must hit $A$ at the first step. That is $h(i,A)=1$. Now equation (\ref{E:lineq}) implies that $h(i,A)\geq 1$
so it follows that $A$ must be an optimal set for its own cardinality. 
$\Box$

Thus the problem of finding an optimal subset is partially resolved if one can construct a vertex cover. Fortunately, there is a simple greedy algorithm (sometimes called the Two Opt algorithm) for constructing a maximal match, whose vertices are a vertex cover.

\begin{figure}[tbp] 
  \centering
  \includegraphics[width=5.02in,height=2.14in,keepaspectratio]{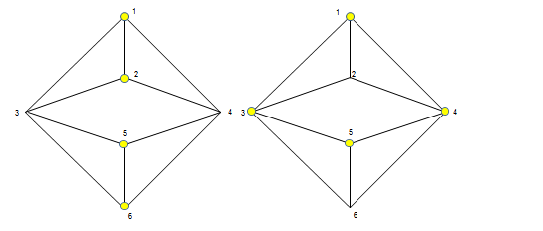}
  \caption{graph for EXAMPLE 2 with 6 nodes is shown with two vertex covers. The nodes of the covers are colored. On the left, the vertex cover contains no optimal subsets except itself, the vertex cover shown on the right contains optimal subsets for $K=1$ through $4$}
  \label{fig:JW6vc-all}
\end{figure}

\begin{Definition}
A \underline{maximal match} of a graph is a set of edges that are non-adjacent (i.e. they do not share a common vertex). The set is maximal in the sense that there is no larger set with this property.
\end{Definition}
As is well known \cite{Jungnickel}, the vertices of a maximal match form a vertex cover. To see why note that
every edge $e$ in $E$ is either an edge of a maximal match or is adjacent to such an edge. Thus $e$ contains a vertex  
in the match. That is, $e$ is incident to some vertex in the match so the definition of vertex cover is satsified. Now let
$\mathcal{M}$ be the set of vertices of the maximal match that was constructed using the Two Opt algorithm. It can be shown (\cite{Cormen}) that,
\begin{equation}
 OPT \leq |\mathcal{M}| \leq 2 \cdot OPT,
\end{equation}
where $OPT$ is the cardinality of the minimal vertex cover. The run time of Two Opt is $\Large{O}(|E|)$ \cite{Cormen}. Because supersets of a vertex cover are optimal sets and since we observe  that optimal sets are often subsets of a vertex cover, it is natural to seek solutions of the optimization problem among the subsets of $\mathcal{M}$:
\begin{equation}
\min_{A \subset \mathcal{M}, \\ |A| \leq M} F(A).
\end{equation}
Recall that  the graph in  EXAMPLE 1 has optimal sets which cannot be found by application of the greedy algorithm. However when the Two Opt algorithm is used to obtain
edges of a maximal match $(1,3),(5,6),(7,8)$, the vertices $\mathcal{M}$, contain optimal 
subsets for $M=1,2,3,4,6$. In contrast to the greedy approach of building up to an optimal set, we start with a maximal match and obtain optimal sets of smaller cardinality as subsets. Unfortunately as the next example illustrates, this approach is not always successful. \\ \\
{\bf EXAMPLE 2:}\\
The vertices of two maximal matches for a graph are shown in Figure \ref{fig:JW6vc-all}. The figure on the left shows a maximal match whose vertex set $\mathcal{M}$ contains no optimal subsets except itself while the match shown on the right contains subsets that are
optimal sets for $M=1$ through $4$. Both maximal matches were obtained using the Two Opt algorithm. \\

Do we have a way to predict when optimal sets of predetermined cardinality are contained in a maximal match?  Presently we do not. The reason is that optimality is not always preserved by adding or removing elements from a single optimal set. However as  discussed in the next section and \ref{S:graph} there is a greedoid containing optimal and near optimal sets (see the definition in \ref{E:Lck}) with a structure that is preserved under such operations. Moreover in section \ref{S:proofs} we prove sufficient conditions for a vertex cover to contain optimal and near optimal subsets (Proposition \ref{P:matchtowatch}).
%
%

\subsection{Optimal and Near Optimal Sets}\label{S:optnopt}
In section 1 a measure of the spread effectiveness of sets was introduced in (\ref{E:hitnumber}). It will be convenient to
convert this to a rank defined on subsets of $V$. In particular, suppose there exists a maximal match with $K$ vertices.
We will order all non-empty subsets $A \subseteq V$ such that $|A|\leq K$ with a ranking function $\rho(A)$ defined as,
\begin{equation}\label {E:rank}
\rho(A)=\frac{F_{max}-F(A)}{F_{max}-F_{min}} 
\end{equation}
where $F_{max}=\max_{\emptyset \ne A\subseteq V| A|\leq K}\large F(A)$, and $F_{min}$ is the corresponding minimum. $F_{min}$ can be calculated by computing $F$ for a maximal match of cardinality $K$, while $F_{max}$ is the maximal value of $F$ among all one element subsets. We assume that $F_{max}\ne F_{min}$. If this were not the case, $F(A)$ would have the same value for any non-empty subset $A$ with $|A| \leq K$. Thus any $A$ would be a solution of the problem.

If $A$ is optimal set of cardinality $K$, then $\rho(A)=1$, the maximum value of $\rho$, conversely the worst performing set has value $0$. An optimal set of size $M <K$, has the largest $\rho$ value among sets of size $M$.
For a constant $c,\, 0<c\leq 1$ and $K$, the non-empty set
\begin{equation}\label{E:Lck}
\large{\mathnormal{L}_{c,K}}=\{A: A\subseteq V, |A|\leq K , \rho(A)\geq c\}
\end{equation}
 defines a set of optimal and near optimal subsets, with the degree of near optimality depending of course on $c$. 
The structure of optimal and near optimal sets is conveniently described in terms of a concept in combinatorial optimization known as a {\it greedoid}  \cite{Korte, Bjorner}.
\begin{Definition}\label{D:greedoid}
Let $\mathbf{E}$ be a set  and let $\mathcal{ F}$ be a collection of subsets of $\mathbf
{E}$. The pair $(\mathbf{E}, \mathcal{F})$  is called a \underline {greedoid} if $\mathcal{F}$ satisfies
\begin{itemize}
  \item $\mathbf{G1}:$ $\emptyset\in \mathcal{F}$ \label{null}
  \item $\mathbf{G2}:$ For $A \in \mathcal{F}$ non-empty, there exists an $a \in A$ such that $A \setminus \{a\} \in \mathcal{F}$ \label{access}
  \item $\mathbf{G3}:$ Given $X$, $Y$ $\in \mathcal{F}$ with $|X| > |Y|$, there exists an $x \in X\setminus Y$, such that $Y \cup \{x\} \in \mathcal{F}$ \label{augment}
\end{itemize}
\end{Definition}
A set in $\mathcal{F}$ is called \underline {feasible.} Note that $\mathbf{G2}$ implies that a  single element can be removed from a feasible set $X$ so that the reduced set is still feasible. By repeating this process the empty set eventually is reached. Conversely
starting from the empty set, $X$ can be built up in steps using the $\mathbf{G3}$ property.\\
Our first step is to show that $\mathnormal{L_{c,K}}$ satisfies condition $\mathbf{G3}$ of the definition
for $0 < c \leq 1$, $0\leq K \leq N$ (Proposition \ref{P:upextend}). The proof depends on several short lemmas. The first uses
an adaptation of an argument in Clark et al.
\begin{lemma}\label{L:mono}
Let $S \subseteq V$, $u \in V\setminus S$. Then $F(S) \geq F(S\cup \{u\})$. 
\end{lemma}
{\bf Proof:} Suppose $S$, a set of nodes is a target set for the random walk. Let $E_{ij}^{l}(S)$ be the event,
$E_{ij}^{l}(S)=\{ X_{0}=i \in V,\   X_{l}=j \in V\setminus S,\ X_{r} \notin S,\ 0 \leq r \leq l \}$.
Thus paths of the random walk start at $i$ and arrive at $j$ without visiting $S$ during the interval $[ 0 , l ]$.
Also define the event $F_{ij}^{l}(S,u)=E_{ij}^{l}(S) \cap \bigcup_{m=0}^{l} \{ X(m)=u\}$ where $u \notin S$. 
Paths in this event also start at $i$ and arrive at $j$ without visiting $S$, but  must visit the element $u$ at some time during the interval $[ 0 , l ]$. Since a path either visits $u$ in the time interval $[ 0 , l ]$ or it does not, it follows that:
\begin{equation}\label{E:seteq}
E_{ij}^{l}(S)=E_{ij}^{l}(S \cup \{u \})\cup F_{ij}^{l}(S,u)
\end{equation}
We have  $E_{ij}^{l}(S \cup \{u\}) \bigcap F_{ij}^{l}(S,u)=\emptyset$. This implies that,
\begin{equation}\label{E:indeq}
\chi(E_{ij}^{l}(S))=\chi(E_{ij}^{l}(S \cup \{u\}))+\chi(F_{ij}^{l}(S,u))
\end{equation}
and therefore:
\begin{equation}\label{E:notindeq}
\chi(E_{ij}^{l}(S)) \geq \chi(E_{ij}^{l}(S \cup \{u\})
\end{equation}
Here $\chi(A)$ is the indicator function of the set $A$. Recalling that $T_{S}$ is the hitting time for
set $S$, the following relation comes from taking the expection of $\chi(E_{ij}^{l}(S))$ on the left hand side of (\ref{E:notindeq}) summing over all $j \in V\setminus S$. Here $\mathbb{E}$ denotes expectation.
\begin{equation}\label{E:hitS}
\mathbf{Prob}\{ T_{S} > l | X_{0}=i\}=\mathbb{E}\left(\sum_{j \in V \setminus S}\chi(E_{ij}^{l}(S))\right)
\end{equation}
A similar result is obtained for $T_{S\cup \{u\}}$ from taking the expectation of $\chi(E_{ij}^{l}(S\cup\{u\}))$ on the right
hand side of (\ref{E:notindeq}) and summing over $j \in V\setminus S$.
Summing once again over all $l \geq 1$ results in the inequality,
\begin{equation}\label{E:hitsumineq}
h(i,S) \geq h(i,S\cup \{u\})
\end{equation}
Finally on summing (\ref{E:hitsumineq}) over all $i$ and recalling the definition of $F$ (equation (\ref{E:hitnumber}))
one obtains the result to be proved.
$\Box$ \\ \\
The following result uses that fact that $F$ is supermodular.
\begin{lemma} \label{L:paving}
For $\bar{c} > 0$, let $\mathcal{P}=\{X \subseteq V \colon F(X) \geq \bar{c}>0 \}$.  If $A, B \in \mathcal{P}$ where $|A|=|B|$ and
$|A\cap B|=|A|-1$, then $A \cap B \in \mathcal{P}$.
\end{lemma}
{\bf Proof:} The hypothesis implies the existence of a set $X$ such that $A=X\cup \{a\}$ and $B=X\cup\{b\}$ with
$a \ne b, a,b, \in V$. The supermodular property of $F$ implies that:
\begin{equation}\label{E:supermod}
F(X\cup \{a\}\cup\{b\})+ F(X) \geq  F(X\cup \{a\})+F(X\cup\{b\})
\end{equation}
Rearranging we have,
\begin{equation}\label{E:step2}
F(X)\geq F(X\cup \{a\})+ [F(X\cup \{b\})-F(X\cup\{a\}\cup\{b\})]
\end{equation}
Thus on writing $X=A\cap B$, using the hypothesis on $A$, and then applying Lemma \ref{L:mono} to the bracketed quantity, we have,
\begin{equation*}
F(A\cap B) \geq F(A)\geq \bar{c}
\end{equation*}
$\Box$. \\
The following lemma is part of a result in \cite{Korte,Bjorner} on paving greedoids.

\begin{lemma} \label{L:upextend}
If $\mathcal{P}$ is any class of sets satisfying the conclusion of Lemma \ref{L:paving}, $\mathcal{S}=2^{V}\setminus \mathcal{P}$
has property $\mathbf{G3}$. That is, given any $A ,\ B \in \mathcal{S}$, with $|A| > |B|$, there is an $a \in A\setminus B$ such that
$B \cup \{a\} \in \mathcal{S}$.
\end{lemma}
{\bf Proof:} Suppose the conclusion is false. If $|A \setminus B|=1$, then for $a \in A\setminus B$, $B \cup \{a\} \in \mathcal{P}$.
But $A=B \cup \{a\}$. To see this suppose there is some $a^{'} \neq a$ that is not in $B$. Then $A \setminus B$ contains $a^{'}$ so $|A\setminus B|>1$ so this is a contradiction. Thus $A=B\cup\{a\}$, but $A \in \mathcal{S}$, and this is also a contradiction.
 Suppose next that $|A \setminus B|>1$. Then there exists $a$, $a^{'}$ 
$\in A \setminus B$. We have $B \cup \{a\}$ and $B \cup \{a^{'}\} \in  \mathcal{P}$. Thus the conclusion of Lemma \ref{L:paving}
implies that $B \in \mathcal{P}$, which is a contradiction.
$\Box$ \\

\begin{proposition}\label{P:upextend}
For $0 < c\leq 1$ and $0 < K \leq N$, let $\mathnormal{L}_{c,K}$ be the class of sets defined in equation (\ref{E:Lck}).
Then $\mathnormal{L}_{c,K}$ satisfies condition  $\mathbf{G3}$.
\end{proposition}
{\bf Proof:} If $\mathcal{P}$ is the set defined in Lemma \ref{L:upextend} then $\mathnormal{L_{c,K}}=\mathcal{S}$ for some $c$.
In fact we may set{ $c=\frac{F_{max}-\bar{c}}{F_{max}-F_{min}}$. If $F_{min}\leq \bar{c} \leq F_{max}$ , we have $0 \leq c \leq 1$.
$\mathnormal{L_{c,K}}$ satisfies the conclusion of Lemma \ref{L:upextend} and therefore it satisfies property $\mathbf{G3}$.
$\Box$  \\ \\
The proposition establishes that $L_{c,K}$ satisfies the $\mathbf{G3}$ property for greedoids. However, $\mathbf{G2}$ does not hold. For example
if the set $A$ has cardinality $m$ where $m$ is the size of the smallest set in $L_{c,K}$ then $A\setminus \{a\}$ cannot be in 
$\mathnormal{L_{c,K}}$  for any element $a \in A$. Conversely, let $c_{m}=\max_{|X| \leq m}\rho(X)$. If $c_{m}\geq c>c_{m-1}$ then $m$ is the size of the smallest set in $\mathnormal{L_{c,K}}$. To create a class of sets with the $\mathbf{G2}$ property, one constructs subsets of $G_m$ of size $n \leq m$ that satisfy $\mathbf{G3}$, while sets $G_n$ for $n > m$ are culled so the remaining sets satisfy $\mathbf{G2}$. The greedoid will then consist of selected subsets and supersets of $G_m$ This construction is illustrated in the next section (section \ref{S:method}) where an example of a greedoid is presented and our proposed method for solving problem (\ref{E:opt}) is demonstrated. It is based on restricting the search for solutions to feasible optimal and near optimal sets in the greedoid.
Following this, in section \ref{S:proofs}, sufficient conditions for the existence of a greedoid are presented as well as proofs.
\subsection{Greedoid Based Approach to the Optimization Problem}\label{S:method}
Let $G$ be the graph discussed in EXAMPLE 1 of section \ref{S:maxmatch}. The smallest sets in $\mathnormal{L_{7/8,8}}$ have cardinality $m=2$.
Any sets in $\mathnormal{L_{7/8,8}}$ will have guaranteed degree of optimality $7/8$ as defined by equation (\ref{E:rank}). Our method is based on searching for optimal and near optimal sets that are the feasible sets of the greedoid constructed from $G_2$.
Values of $F$ for two element sets are computed in advance to start the procedure.
In addition to $G_2$  itself, the empty set and selected one element subsets of $G_2$ must also be included. These sets arise from the pairwise intersection of sets in $G_2$ (listed below) and in addition they must satisfy $\mathbf{G3}$ for every set in $G_2$.
The two element sets are:
\begin{center}
Elements of $G_{2}=\widehat{G}_2$
\begin{align} \label{E:twoelement} 
& \{3,7\}^{*} , \{2,8\}   \\    \nonumber
& \{1,7\}^{*} , \{1,9\}    \\   \nonumber
&  \{2,7\}    , \{2,9\}     \\    \nonumber
& \{3,8\}^{*} , \{5,7\}^{*}  \\    \nonumber
& \{3,9\}     , \{1,4\}       \\     \nonumber
& \{3,4\}     , \{1,6\}^{*}    \\      \nonumber
& \{3,6\}^{*} , \{2,4\}       \\     \nonumber
& \{1,8\}^{*} , \{2,6\}        \\      \nonumber
\end{align}
\end{center}
The one element sets of the greedoid are therefore  $\widehat{G}_{1}=\{1,2,3,7\}$. The two element feasible sets are defined to be 
supersets of $\widehat{G}_{1}$. Here that is all of $G_2$, no culling is necessary. Once $n >m$, the feasible sets of the greedoid are subsets of $G_n$ that are supersets of $\widehat{G}_{n-1}$. Thus feasible sets of cardinality $3$, $\widehat{G}_{3}$ are sets in $G_3$ that are 
supersets of $\widehat{G}_2$. It will be helpful in our discussion of the method to list some of the elements in the greedoid.
It can be checked that the displayed sets satisfy $\mathbf{G1}$-$\mathbf{G3}$.
\begin{align}  \label{E:morelements} 
 &\text{10 elements of }\widehat{G}_3 & &\text{10 elements of } \widehat{G}_{4}&&\text{10 elements of }\widehat{G}_5    \\   \nonumber
 &\{3,4,9\}    &  &\{2,5,7,9\}    & & \{2,3,4,6,9\}  \\   \nonumber 
 &\{3,6,8\}    &  & \{2,5,7,8\}   & & \{2,3,4,6,8\}   \\   \nonumber
 &\{1,5,7\}    &  & \{1,5,7,9\}   & & \{1,3,4,6,9\}    \\   \nonumber
 &\{2,5,7\}    &  & \{1,5,7,8\}   & & \{1,3,4,6,8\}    \\    \nonumber
 &\{3,4,7\}    &  & \{3,4,6,9\}   & &\{2,3,6,7,8\}    \\    \nonumber
 &\{3,5,7\}    &  & \{3,4,6,8\}   & & \{2,3,5,7,9\}     \\   \nonumber
 &\{3,7,9\}    &  & \{3,6,7,8\}   & &  \{2,3,5,7,8\}     \\   \nonumber
 &\{3,6,7\}    &  &\{3,5,7,9\}    & & \{2,3,4,7,9\}     \\    \nonumber
 &\{3,7,8\}    &  & \{3,5,7,8\}   & & \{2,3,4,6,7\}     \\    \nonumber
 &\{3,4,6\}    &  & \{3,4,7,9\}   & & \{1,3,6,7,8\}     \\    \nonumber
\end{align}
Suppose we try to solve the optimization problem (\ref{E:opt}) for $M=5$. One could exhaustively compare the $F$ values for all possible $5$ element subsets. Alternatively, one could apply the greedy algorithm to the best single node(s).The thesis of this paper is that one can reduce the number of $5$ element sets to the smaller class of near optimal sets. For any $c>0$ this class contains optimal $5$ element sets and approximations that are better than the guarantees of the greedy algorithm approximations. The greedoid properties enable the near optimal sets to be constructed from one-element sets in a stepwise manner.
Given the vertex cover for our example graph $G$, $VC=\{1,3,5,6,7,8\}$ , a selection of two element sets in $\widehat{G_2}$  that are also subsets of $VC$ (indicated in display (\ref{E:twoelement}) by stars) was used to create a sample of $5$ element candidate sets to be compared. Each starred $2$ element set was greedily extended to a $5$ element candidate set, step by step. The left most column of display (\ref{E:best3}) shows the $3$ element sets obtained from a greedy one element extension of the $2$ element sets. The next column shows the corresponding $F$ values of each set. The third column is a list of the $4$ element sets obtained by a one element greedy extension of the $3$ element sets. The column of corresonding $F$ values appears next. Finally the fifth and sixth columns contain the $5$ element sets and the corresponding $F$ values respectively. 

Our method is to offer the $5$ element set with smallest $F$ value as the approximate(or actual) solution of the problem.In this case the set we obtain, $A=\{2,3,4,6,8\}$ is a solution. Note that the same procedure can be used to obtain the solutions to the problem for $M=3$ ($\{3,6,8\}$) and $M=4$ ($\{2,5,7,8\}$ or $\{1,5,7,9\}$). The set $\{1 ,5,7,8\}$ is symmetric to the latter sets so it is left out.

This example illustrates two more additional advantages of the greedoid structure of optimal and near optimal sets. Optimal solutions that are not subsets of the vertex cover but that are in the greedoid can still be found by these methods. 
For example, $A$ is not a subset of $VC$. Secondly, an approximate solution can be improved using greedoid properties $\mathbf{G2}$ and $\mathbf{G3}$. Thus given an approximation $B=\{2,3,4,6,7\}$ with $F(B)=5.00$, the sequence $\{2,3,4,6,7\} \rightarrow \{3,4,6,7\} \rightarrow \{3,4,6\} \rightarrow \{3,4,6,8\}\rightarrow A= \{2,3,4,6,8\}$.Thus swapping the elements $7$ in $B$ with $8$ in $A$ preserves near optimality and in fact improves the value of $F$. Thus we conjecture that navigation through the greedoid offers a systematic way of accessing highly optimal solutions. Section \ref{S:graph} elaborates on this idea. 
\begin{align}\label{E:best3}
&\{3,4,7\}  & &  12.4 && \{3,4,7,9\}&& 8.00 && \{1,3,4,7,9\}&& 5.00  \\  \nonumber
&\{3,6,7\}  & &  12.4 && \{3,4,6,7\}&& 8.00 && \{2,3,4,6,7\}&& 5.00  \\  \nonumber
&\{1,5,7\}  & &  11.8 && \{1,5,7,9\}&& 7.40 && \{1,2,5,7,9\}&& 5.00  \\  \nonumber
&\{2,5,7\}  & &  11.8 && \{2,5,7,8\}&& 7.40 && \{1,2,5,7,8\}&& 5.00   \\  \nonumber
&\{3,6,8\}  & &  11.6 && \{3,4,6,8\}&& 7.82 && \{2,3,4,6,8\}&& 4.82 \\  \nonumber
&\{1,4,6\}  & &  13.9 &&  \{1,4,6,9\}&&9.04 && \{1,4,5,6,9\}&& 6.21   \\  \nonumber
\end{align}
\section{The Structure of Optimal and Near Optimal Target Sets} \label{S:proofs}
In this section we discuss how given a fixed $c$, a measure of near optimality and $K$ the size of the largest optimal set
under consideration, a greedoid can be constructed from the set $L_{c,K}$ of optimal and near optimal sets.  Two sufficient conditions are presented (see Case I and Case II) for the construction of a greedoid. Corollaries 1 and 2 describe the feasible sets of the greedoid for Cases I and II respectively.

In order to construct a greedoid of optimal and near optimal sets one must construct a class of sets from $L_{c , K}$ that satisfy $\mathbf{G2}$ as well as $\mathbf{G3}$. The process can proceed along two tracks- one for sets of cardinality $n\leq m$ and the second for sets of cardinality $n \geq m$.  For the latter, let $G_{n}=\{ A \in L_{c, K} ,\  |A|=n \}$ for $n \geq m$. Lemma \ref{L:upextend}
shows that $\mathnormal{L}_{c,K}$ satisfies $\mathbf{G3}$.
We set $G_{m}=g_{m}$. Suppose $\widehat{G}_{n}$, for $m\leq n\leq K$,( respectively $\{\widehat{g}_n\}_{1 \leq n \leq m}$ )  is a class of  supersets (respectively subsets) of $\widehat{G}_m$ of cardinality $n$, with the properties:

\begin{description}
\item[T1] $\widehat{g}_m=\widehat{G}_{m} \subseteq G_m$ 
\item [T2] Every $A \in \widehat{g}_n$, contains a subset $B \in \widehat{g}_{n-1}$.
\item[T3] For  every $B \in \widehat{g}_{n-1}$,  and $A\in \widehat{g}_{n}$, there is 
          $p \in  A\setminus B$ such that  $B \cup \{p\} \in \widehat{g}_{n}$.
\item[T4] For $n > m$, $\widehat{G}_{n}=\{A|A\in G_{n}\ A\supset B, \  B\in \widehat{G}_{n-1}\}$,                   
\end{description}
then the collection of sets $\mathcal{F}_{c,K}=\{\,\emptyset,\ (\widehat{g}_{n},\ 1 \leq n \leq m) , \, (\widehat{G}_{n}, 
m \leq n \leq K),\}$ are the feasible sets for a greedoid over ground set $V$.
This follows from Lemma \ref{L:T-ax} and its consequence Theorem \ref{T:structure}.
\begin{lemma} \label{L:T-ax}: Suppose $\widehat{G}_{n}$, $n \geq m$,(respectively  $\widehat{g}_n$ $n < m$ ) are supersets (respectively subsets) of $G_m$ that satisfy $\mathbf{T1}$-$\mathbf{T4}$. Then: 
\begin{description}
\item [(1)]$f_m=\{ \,\emptyset , \ (\widehat{g}_{n}, \ 1 \leq n \leq m) \ \}$ satisfies $\mathbf{G1}$ and $\mathbf{G2}$.
\item[(2)] If $A \in \widehat{G}_{n}$ for $n >m$, there is an $a \in A$ such that $A\setminus \{a\}$ is in $\widehat{G}_{n-1}$.
\item[(3)] If $B \in \widehat{g}_n , n < m$ and $A \in f_m$ or $\in G_{k}, k\geq m$, with $|A| > |B|$, then there is an $a \in A$            such that $B \cup \{a\} \in \widehat{g}_{n+1}$.
\end{description}
\end{lemma}
Proof: For statement (1) $\mathbf{G1}$ is clear.  By $\mathbf{T2}$ for any $A \in \widehat{g}_n$  there is a subset $B \in \widehat{g}_{n-1}$. Since $A$ has cardinality $n$ and $B$ has cardinality $n-1$ we must have $B=A\setminus \{a\}$ for some $a$. 
Thus $\mathbf{G2}$ holds. The proof of statement(2) is the same as the proof of the $\mathbf{G2}$ property for $f_m$ where here $\mathbf{T4}$ is used.
  To show (3), first suppose $A \in \widehat{g}_{k}, k >n$.  By (1) we may apply  $\mathbf{G2}$ 
repeatedly to  reduce $A$ to  a set $A^{'} \in \widehat{g}_{n+1}$. Then by $\mathbf{T3}$, $B \cup \{a\} \in \widehat{g}_{n+1}$ for some $a \in A^{'}\setminus B$. Since $A^{'} \subseteq A$, we have $a \in A$. Next if $A \in \widehat{G}_{k}$ for $k\geq m$, either $A \in \widehat{G}_{m}$ or (2) can be applied repeatedly to produce a set in $\hat{G}_{m}$. By $\mathbf{T1}$, $\widehat{G}_{m}=\widehat{g}_{m}$ so   $A$ is reduced finally to a set in $f_m$.  Thus we obtain the conclusion by repeating the argument used in the previous case. 
$\Box$. \\
\begin{theorem} \label{T:structure} The class of sets $\mathcal{F}_{c,K}=\{\,\emptyset,\ (\widehat{g}_{n},\ 1 \leq n \leq m-1) , \, (\widehat{G}_{n}, m \leq n \leq K) \}$  is a class of feasible sets for a greedoid over $V$.
\end{theorem}

Proof: By Proposition \ref{P:upextend}, $\mathbf{G3}$ is satisfied when $A \in \widehat{G_n}$ since 
$\widehat{G_n}\subseteq G_{n}$. $\mathbf{G3}$ is established for $A \in \widehat{g_k}$ by Lemma \ref{L:T-ax} (3).  Property $\mathbf{G2}$ for $A \in \widehat{G}_{n}$ follows from Lemma \ref{L:T-ax} (2) when $n > m$ and Lemma \ref{L:T-ax} (1) and $\mathbf{T1}$ when $A \in \widehat{G}_{m}=\widehat{g}_{m}$. If  $A \in \widehat{g}_{k}$, then $A$ satisfies $\mathbf{G2}$ because of Lemma \ref{L:T-ax} (1). Finally $\mathcal{F}_{c,K}$ clearly contains $\emptyset$.
$\Box$.\\  \\
Theorem \ref{T:structure} describes the feasible sets of the greedoid formed by sets $\widehat{g}_{n}$ and $\widehat{G}_n$ when
$\mathbf{T1-T4}$ are satisfied.
We next present two sufficient conditions for the existence of sets satisfying $\mathbf{T1}$-$\mathbf{T4}$.\\ \\
Case I:  Suppose $G_m$ is a single set $G_{m}=\widehat{G}_{m}=\{H\}$. For $n=m-1$, define $\hat{g}_{m-1}$ to be the class of subsets of $H$ of cardinality $m-1$. When $n < m-1$, $\widehat{g}_{n-1}=\{B  | B \subset H , |B|=n-1 \}$. $\widehat{G}_n$ is a superset of $G_m$ obtained by stepwise addition of elements  as described in $\mathbf{T4}$. If $G_m$ has more than one set an arbitrary $H$ can be selected.

\begin{corollary} \label{C:case1} The collection of sets $\{\widehat{g}_{n}| n\leq m \}$ and $\{\widehat{G}_{n} | n>m \}$ in Case I,
satisfy  conditions $\mathbf{T1-T4}$, therefore $\mathcal{F}(L_{c,K})=\{\emptyset, \ (B: B \subset H, |B|<m ),\ H, \ (A| A\supset H , |A| \leq K )\}$ are the feasible sets of a greedoid.
\end{corollary}}

Proof: $\mathbf{T1}$ and $\mathbf{T2}$ follow directly from the definitions of $\widehat{G}_n$ and $\widehat{g}_{n}$. The property $\mathbf{T3}$ holds. To see this suppose $A \in \widehat{g}_n$ and $B \in \widehat{g}_{n-1}$. Since $A$ and $B$ are subsets of $H$, $A$ has at least one element $p$ that is not in $B$. Now $\hat{g}_{n}$ contains all subsets of $H$ of cardinality $n$ it must have $B \cup \{p\}$. $\mathbf{T4}$ follows immediately from the definition of $\widehat{G}_n$.
$\Box$.\\ \\ 
To describe the second sufficient condition for the existence and construction of a greedoid we will need a couple of definitions.\\ \\
{\bf Definition:} Given subsets $A$ and $B$ of $V$, with an element $p \in V\setminus B$ such that $A=B \cup \{p\}$, $A$ is a \underline{parent}  of $B$ and $p$ is a \underline{partner} of $B$. \\

As before members of $G_n$ are elements of $L_{c,K}$ of cardinality $n \geq m$ and $g_n$ are subsets of $G_m$ of 
cardinality $n\leq m$. However in the present situation, sets in $g_n$ will be defined in terms of pairwise intersections of sets of cardinality $n+1$.  Specifically let $U_{n}=\{ B: B=C \cap A, \, |A|=|C|=n+1 \}$. Then for $n<m$ we define by backward induction
starting from $m$,
\begin{equation*}
g_{n}= \{B \in U_{n}| B=E \cap F, \, E, F \in g_{n+1}\}. \\
\end{equation*}
Thus elements in $g_n$  subsets of size $n$ are pairwise intersections of adjacent pairs (in the Hamming metric sense) of sets in $g_{n+1}$. \\

 The second sufficient condition is defined in terms of the following sets:
let $X_{p,k}=\{A | A \in g_{k} , A=S \cup \{p\}, |S|=k-1\}$ , $Y_{p,k}=\{W | W \in g_{k-1}, W \cup \{p\} \in g_{k} \}$.
The set $A$ is in $X_{p,k}$ if it contains $p$ and is the parent of an $S$ with cardinality $k-1$. A set $W$ is  in $Y_{p,k}$  if it has a partner $p$. \\ \\
\begin{proposition} \label{P:case2}
 Suppose $B \in g_{n}$ where $n < m$. Further suppose there is a finite set $\{p_{i} \in V: i=1 \cdots  l_{n}\}$ (which may depend on $B$), such that $g_{n+1}=\bigcup_{i=1}^{l_n}X_{p_{i},n+1}$ where $p_{i} \notin B, \ i=1 \cdots, l_{n}$,
and $B \subset \bigcap_{i=1}^{l_n}Y_{p,,n+1} \neq  \emptyset$. Then for every  $A\in g_{n+1}$, there exists a $p \in A\setminus B$ such that $B \cup \{p\} \in g_{n+1}$. \\
\end{proposition} 
{\bf Proof:} The hypothesis states that $B$ has partners, $p_{i} \notin B, i=1,\cdots l_{n}$. Moreover we also have that
each $A \in g_{n+1}$ contains an element, say $p$ in this set by hypothesis. Thus by the definition of partner,
we must have $B \cup \{p\} \in g_{n+1}$.
$\Box$\\
The shorthand notation $B \nearrow g_{n+1}$ used in the sequel means that $B$ satisfies the hypothesis of Proposition \ref{P:case2}. Therefore as a consequence of the conclusion, $B$  satisfies $\mathbf{T3}$. \\ \\
Case II

\begin{itemize}
  \item (i) For every $1\leq n \leq m$, $\widehat{g}_{n}=\{ B \in g_{n}\mid B \nearrow g_{n+1}, B\supset C, C\in \widehat{g}_{n} \}\ne \emptyset$.
   \item     (ii)   $\widehat{g}_{m}=\widehat{G}_{m}=\{A \in G_m | A \supset B, B \in \widehat{g}_{m-1}\}$.
    \item    (iii)  $\widehat{G}_{n}=\{A \in G_n | A \supset B \in G_{n-1} \}$ , $n>m$. \\ \\
 \end{itemize}
{\bf REMARK:} If property (i) is true then for every $n < m$, there are elements in $g_{n}$ that satisfy the hypotheses of Proposition 
\ref{P:case2}. \\
Case II is illustrated in the example discussed in section 2.3. Here the one element feasible sets are
are $ \widehat{g_1}=\{\{1\},\{2\},\{3\},\{7\}\}$. 
The elements $\{4\}$,$\{5\}$,$\{6\}$,$\{8\}$,$\{9\}$ are excluded even though these sets arise from the pairwise intersection of sets in $G_2$, because they fail to satisfy (i) and therefore they do not satisfy $\mathbf{T3}$ (and thus $\mathbf{G3}$).

\begin{proposition}\label{P:case2-T} If CASE II holds, then $\mathbf{T1}$-$\mathbf{T4}$ is satisfied.
\end{proposition}
{\bf Proof:}  $\mathbf{T1}$ and $\mathbf{T2}$ follow easily from (i) and (ii). To see that $\mathbf{T3}$ holds note that it is a consequence of (i) since $B \nearrow \widehat{g}_{n+1}$ and any $A \in \widehat{g}_{n}$ is in $g_n$. $\mathbf{T4}$ follows from (iii). $\Box$. \\ 
\begin{corollary}\label{C:case2}
Suppose $\{\widehat{g}_n\}$ $1 \leq n \leq m$ and $\{\widehat{G}_{n}\}$  $m \leq n \leq K$ satisfy the conditions of Case II.Then
the class of sets $\mathcal{F}(L_{c,K})=
\{\emptyset , \  (\widehat{g}_{n} ,\ 1 \leq n \leq m-1),\  (\widehat{G}_{n},\ m\leq n\leq K) \} $
are the feasible sets of a greedoid.
\end{corollary} 
{\bf Proof:} By Proposition \ref{P:case2-T}, the hypotheses of Theorem \ref{T:structure} are satisfied. Thus the conclusion of this proposition follows from the theorem. $\Box$. \\

 When $\mathcal{M}$ is a vertex cover (e.g. the vertices of a maximal match) then
we can give a partial answer to the question raised in section \ref{S:newsolution} of when vertex covers contain optimal sets. If $\mathcal{M}$ is a feasible set with $|\mathcal{M}| > m$, then it will contain optimal or near optimal sets where the degree of optimality is
defined by $c$ in (\ref{E:Lck}). \\ 
\begin{proposition}\label{P:matchtowatch}:
Let $\mathcal{M}$ be the vertices of a vertex cover (maximal match) and suppose $c$ and $K$ as in (\ref{E:Lck}) are given.
Further let $m$ be the minimum cardinality of sets in $\mathnormal{L}_{c,K}$. If $|\mathcal{M}|>m$, and $\mathcal{M}$ is a 
feasible set of the greedoid in Theorem \ref{T:structure}, then it and its subsets 
with cardinality at least $m$ are in $\mathnormal{L}_{c,K}$.Thus it has nearly optimal subsets. 
In particular if there is an $S \in \widehat{G}_m$ with $S \subset \mathcal{M}$,
then $\mathcal{M}$ has nearly optimal subsets in the sense of (\ref{E:Lck}). \\
\end{proposition}
When are any of the non-optimal sets contained in $\mathcal{M}$ actually optimal? In general we do not know. Since $c$ effectively measures the quality of the sets in $L_{c,K}$ the closer $c$ is to $1$, the closer the subsets are to optimal sets. A step towards answering this question would be to identify a  class of graphs for which a moderate level of $c$ is enough to guarantee that a large percentage of $L_{c,K}$ consists of optimal and very high quality sets.
\subsection{The Graph of Optimal and Near Optimal Sets}\label{S:graph}
We introduce a graph
 $\mathcal{G}(\mathnormal{L}_{c,K})$ whose nodes are the feasible sets of the greedoid described in Theorem \ref{T:structure}. 
To simplify the notation we  use $\mathsf{F}=\mathcal{F}(\mathnormal{L_{c,K}})$ to denote the node set of 
$\mathcal{G}(\mathnormal{L}_{c,K})$. The local structure of the graph is defined by adjacent nodes.
\begin{Definition}\label{D:adjacent}
Two nodes $A$ and $B$$\in \mathsf{F}$ are adjacent in $\mathcal{G}(\mathnormal{L}_{c,K})$ if one of the following
statements is true.
\begin{itemize}
  \item $A$, $B$ $\in \mathsf{F}$, $B=A \cup \{r\}$, for $r \notin A$
  \item $B=A\setminus a$ for some element $a \in A$
  \item $|A|=|B|$ and $|A\setminus B|=1$
\end{itemize}  
\end{Definition}
Let $A$ and $C$ in $\mathsf{F}\cap\mathnormal{L_{c,K}}$ be two feasible sets of equal cardinality. We assume the cardinality
is greater than $m$,the smallest set in $L_{c,K}$. For some $a \in V$, the set
$D=A \setminus a$ is $\in \mathnormal{L}_{c,K}$. Indeed, since  $A$ is feasible, for some $a$, $D$ is feasible.
Moreover $|D|\geq m$. By Theorem \ref{T:structure}, any set of this cardinality is a member of $\mathnormal{L}_{c,K}$.
Feasible sets of cardinality greater than $m$ that are described in Corollaries \ref{C:case1}
and \ref{C:case2} have the property that there is a $d \in C$, not in $A$, such that $B=D \cup \{d\} \in \mathsf{F}$.
For such greedoids, sets $A$ and $B$ and $D$ are adjacent
in $\mathcal{G}(\mathnormal{F}_{c,K})$ where clearly $B$ is the result of replacing $a$ by $d$ in $A$. By repeated swapping and other local moves,
one can construct a neighborhood of $A$ suitable for local search. When $A$ is a subset of vertices of a feasible maximal match or vertex cover, navigation to an enlarged neighborhood can be achieved by a sequence of moves to feasible adjacent sets. In fact optimal sets that are not subsets of the vertex cover can be reached. In section \ref{S:method}, this is demonstrated using a different type of
path than the one discussed here.
A topic for future research is the development of efficient methods for doing this calculation as well as navigating $\mathcal{G}(\mathnormal{L_{c,K}})$ so that the number of evaluations of $F$ is minimized.
\section{Conclusion}\label{S:concl} 
We posed the problem of identifying the subset of nodes in a network that will enable the fastest spread of information in a  decentralized communication environment. In a model of communication based on a random walk on an undirected graph $G=(V,E)$, the optimal set of nodes are found by minimizing the sum of the mean times of first arrival to the set by walkers who start at nodes outside the set.

Since the objective function for this problem is supermodular, the greedy algorithm has been a principal method for constructing approximations to optimal sets. References \cite{Clark}, \cite{Borkar} obtain results guaranteeing that these sets are in some sense within 
$(1-1/e)$ of optimality.
   In this work we took a different approach. Rather than seek an optimizing set for problem (\ref{E:opt}) without any information
about $G$ other than its cardinality and objective function $F$-- the problem was reformulated.

We introduced the concept of optimal and near optimal set, ordering the feasible subsets of problem (4) with a ranking relative to the vertex cover of the graph  with cardinality $K$. A constant $c$ is a lower bound on the rank that measures the degree of optimality of the sets (see equation (\ref{E:Lck})). For a fixed cardinality, higher ranking subsets are close to optimal solutions of the problem and approximations to the problem can be compared. If we want to improve an approximation we need a set structure to enable us to make local moves from one set to another. In particular, it should be possible to  add, delete or swap elements  so that the resulting set  is still optimal or near optimal with the specified degree of optimality $c$.
Our main result (Theorem \ref{T:structure}, section \ref{S:proofs}), describes sufficient conditions for the construction of a greedoid based on selected optimal and near optimal subsets of smallest cardinality. This greedoid provides the desired structure for local search methods. A greedoid graph formalizes this idea as described in section \ref{S:graph}. We believe that a local search method based on branch and bound is a promising avenue for current and future research (\cite{HuntB&B}). As seen in section \ref{S:method} there are graphs where it is possible to improve the bounds guaranteed by the use of the greedy algorithm. As a corollary of our work we prove a theorem that gives sufficient conditions for when a vertex cover contains a near optimal subset, thereby partially confirming empirical observations made in section \ref{S:maxmatch}.

Other issues for additional future research are characterizing the class of graphs for which this approach works well, i.e. when can sets with a high degree of optimality in the sense of problem (\ref{E:opt}) be achieved when we have a large $c$ ? Finally the methods of this paper could be used to optimize submodular, monotone functions that arise in other models of network spread. For example it would be interesting to consider the independent cascade model discussed by Kempe et al. (\cite{Kempe}).



%
%
%
%
%
%

%
%
\end{document}